\documentstyle[12pt]{article}

\begin{document}
\title{Nuclear Matter with Quark-Meson Coupling II:\\
Modeling a Soliton Liquid}
\author{
Nir Barnea\footnote{e-mail: barnea@ect.it}
 and Timothy S. Walhout\footnote{e-mail: walhout@ect.it}\\
{\small\it ECT*, European Centre for Theoretical Studies in Nuclear Physics 
and Related Areas,}\\
\vspace {-0.6mm}
{\small\it Strada delle Tabarelle 286, I-38050 Villazzano (Trento), Italy}\\
{\small and}\\
{\small\it Istituto Nazionale di Fisica Nucleare, Gruppo
collegato di Trento}\\
        }
\maketitle

\setcounter{page}{1}

\maketitle
\begin{abstract}
We study in further detail a nontopological soliton model with
coupling between quarks and mesons selected as promising in
a previous study that employed the so-called Wigner-Seitz
approximation for dense systems. Here we go
beyond this approximation by introducing the disorder 
necessary to reproduce the liquid state, using the
significant structure theory of Jhon and Eyring.
We study nuclear matter, with particular
interest in the transition to a quark plasma.
The model studied is a variation of the chromodielectric
model of Fai, Perry and Wilets, where explicit coupling
to a scalar meson field is introduced.
\end{abstract}

\setlength{\parskip}{0.0in}



\setlength{\textheight}{9.0in}
\setlength{\parskip}{0.1in}
\newpage


\section{Introduction}

\hspace* {6mm}

This is the second of two papers that seek to select and
develop a soliton model of nucleons for application to
dense matter and, in particular, the transition to 
deconfined quark matter. In the previous paper, hereafter
referred to as (I), we presented the motivation for
our choice of a particular
class of nontopological soliton models, the Friedberg-Lee type
models \cite{Fri77}, which have explicit quark degrees
of freedom and a dynamical confinement mechanism resulting
from a composite scalar gluon field that forms a solitonic bag 
in which constituent quarks then reside.
We considered extensions of these models that include
explicit meson degrees of freedom coupled linearly to
the quarks in order to obtain a
reasonable description of nuclear
interactions. For simplicity, we have studied ---
and here also study --- only models
that include scalar and vector mesons, neglecting in
these initial investigations
any possible explicit effects of pions in dense matter (although
the scalar meson can be considered an effective two-pion resonance).
We avoid any double counting of hadronic degrees of
freedom by including only nuclear
constituent quarks in our calculations. Gluons enter
our calculations only through the scalar glueball
field; perturbative gluonic effects are ignored in nuclear matter.

The distinguishing feature between the various models
we considered in (I) is the precise form of the coupling
between the quarks and the glueball field. We compared
the various FL models by studying dense matter within
the Wigner-Seitz approximation. In general, one can
find parameter choices that produce reasonable results
for free nucleon properties independent of the particular
form of the quark-glueball coupling. It is in dense matter,
where quark bags begin to touch, that the various
models are distinguished. We found that models which
have a quark-glueball coupling in accord with the
dictates of the chiral chromodielectric model ($\chi$CD) show
a behavior more in line with phenomenology. In these
models, the quark-gluon coupling is of leading order
two or greater in the glueball field, which is essential
for the elimination of transitions to unphysical quark
plasma phases at unrealistically low densities.
Furthermore, it was found that coupling the quarks
to a scalar meson field ensures saturation.  The quark-meson
coupling is taken to be independent of the glueball field
within the mean field approximation
to avoid unphysical transitions in dense matter, as was
detailed in (I).

Here we wish to further study the model selected in (I)
by going beyond the relatively rough approximations
used there in modeling the liquid state.
The Wigner-Seitz approximation
consists in assuming that each nucleon is confined by
interactions with its nearest neighbors to
a given volume, equal to the inverse of the baryon density,
known as the Wigner-Seitz cell. For a solid cubic lattice,
for example, the Wigner-Seitz cell is a cube. Here, we
are interested not in the solid but rather the liquid
state, and so the usual choice is to take the Wigner-Seitz
cell to be a sphere in the hope that one thereby better
models the disorder of a liquid. This, however, is
clearly not enough: if we look at models of the
liquid state used by physical chemists in order to
describe molecular liquids (models that passed out of
use several decades ago after the development of
large-scale computers enabling the use of
molecular dynamics and Monte Carlo techniques), the
Wigner-Seitz approximation employed in (I) corresponds
to the cell model of Lennard-Jones and Devonshire \cite{LJD37}, 
which does much better at reproducing the solid state than
the liquid state \cite{Bar63}. Instead, we shall employ a refinement
of the cell model --- namely, the Significant Structure
Theory of Jhon and Eyring \cite{JE69} --- which introduces holes
into the system in order to account for the disorder
present in liquids.

As a matter of fact, in (I) we assumed
that the Wigner-Seitz cell is simply a sort of ``average
snapshot'' of the nuclear medium felt by
the quarks inside an otherwise freely moving nucleon. 
Thus we proceeded by
subtracting away energy due to spurious center of mass motion
(due to the fact that the nucleon was not constructed by
putting the quarks in a good momentum state), and then took
the kinetic energy of the system to be that of a free Fermi
gas. Clearly, this approximation can only be justified at
low densities. As the density increases the motion of
an individual nucleon is affected by the medium,
and this leads us to consider the Wigner-Seitz cell not
just as a boundary upon the quark wave functions that
build up a nucleon, but also as a restriction upon the
motion of the nucleon itself. This leads to the considerations
of the previous paragraph, which shall be further developed
in Sec. \ref{sec:liquid}.

The outline of the paper is as follows. The nontopological
soltion model used is reviewed in Sec. \ref{sec:model}.
In Sec. \ref{sec:liquid} we discuss various attempts to
model a soliton liquid, then motivate and introduce the
particular model based on siginificant liquid
structures that we shall use here. 
In Sec. \ref{sec:results} we present the resulting
equations of state for
nuclear matter and discuss the transition
to quark matter. 
A general summary and discussion of the two papers
is given in Sec. \ref{sec:con}.

\section{The Model} 
\label{sec:model}

The nontopological soliton model we study here is based upon
the chiral chromo\-dielectric model
of Fai, Perry and Wilets \cite{Fai88}. In its full version,
the model contains quark and gluon degrees of freedom.
A scalar glueball field $\sigma$ couples to the quarks,
and colored gluons $A_\mu^a$are treated perturbatively. The scalar
field provides absolute confinement of both quarks and
gluons and gives consituents a mass. Meson exchange is
surely present in this model, but for simplicity we alter
the original $\chi$CD by dropping the gluon field $A_\mu^a$ 
and ignoring sea quarks. 
Instead, as in quark-meson coupling 
models, we introduce a scalar meson $\phi$.
The vector meson $V_{\mu}$, which provides repulsion
in quantum hadrodynamics, is not necessary here since the
soliton structure provides repulsion between nucleons, and
so for simplicity we set $V_{\mu}=0$.
We assume the scalar meson couples linearly
to the quarks and take
the quark-meson vertex to be independent of $\sigma$. 
The Lagrangian density for our model is
\begin{eqnarray} \label{Lagrange}
{\cal L} & = &\bar{\psi} \left[ i\gamma^{\mu} \partial_{\mu}
        - g(\sigma) - g_s \phi 
          \right] \psi 
        + \frac{1}{2}\partial_{\mu}\sigma\partial^{\mu}\sigma - U(\sigma) 
        \nonumber \\ & &	
        + \frac{1}{2}\partial_{\mu}\phi\partial^{\mu}\phi 
        - \frac{1}{2}m_{s}^2 \phi^2 
        - \frac{1}{4}F_{\mu\nu}F^{\mu\nu}, 
\end{eqnarray}
where 
\begin{equation} \label{U(s)}
  U(\sigma) = \frac{a}{2!}\sigma^2
            + \frac{b}{3!}\sigma^3
            + \frac{c}{4!}\sigma^4 + B\; \; .
\end{equation}
We have set the current quark masses to zero. The
parameters are as chosen in (I).
The constants of the potential are $a=50$fm$^{-2}$, 
$b=-1300$fm$^{-1}$ and $c=10^4$,  
so that $U(\sigma)$ has a
local minimum at $\sigma=0$ 
and a global minimum at $\sigma=\sigma_v=0.285$fm$^{-1}$,
the vacuum value. The mass of the glueball excitation
associated with the $\sigma$ field
is $m_{GB}=\sqrt{U''(\sigma_v)}=1.82$GeV and the
bag constant is $B=46.6$MeV/fm$^3$.
The quark-$\sigma$ coupling is
\begin{equation} \label{g(s)} 
  g(\sigma) =  g_{\sigma}\sigma_v [\frac{1}{\kappa(\sigma)}-1] ,
\end{equation}
where we choose
the chromodielectric function $\kappa(\sigma)$ to be
\begin{equation} \label{kappa(s)}
 \kappa(\sigma)=1+
\theta(x)x^3[3x-4+\kappa_v]
  \; \; ; \; x=\sigma/\sigma_v \; ,
\end{equation}
In the following we take $g_\sigma=3$ and $\kappa_v=.1$.

We solve the Euler-Lagrange equations in the mean field
approximation, replacing the glueball field $\sigma$
by the classical soliton solution $\sigma(\vec{r})$
and the meson field by its expectation value in the 
nuclear medium $< \phi > = \phi_0 $.
The resulting equations for the quark and the scalar soliton field are
solved in a Wigner-Seitz cell of radius $R$ by implementing
boundary conditions based upon Bloch's theorem, as detailed
in (I). 
The quark spinor in the lowest
band is assumed to be an {\it s}-state 
\begin{equation} \label{q_spinor}
  \psi_k = \left( \begin{array} {c}
	          u_k(r) \\ i \sigma \cdot \hat{r} \; v_k(r)
                  \end{array} 
           \right) \chi,
\end{equation}
and we make the simplifying assumption of
identifying the bottom of the
lowest band by the demand that the derivative of the upper component of the
Dirac function disappears at $R$, and the top of that band by the demand 
that the value of the upper component is zero at $R$ \cite{Bir88}.
The resulting equations for the spinor components are
\begin{equation} \label{u_r}
 \frac{d u_k}{d r} + \left[ g(\sigma) - g_s \phi_0 +
                    \epsilon_k \right] v_k = 0
\end{equation}
\begin{equation} \label{v_r}
 \frac{d v_k}{d r} + \frac{2 v_k}{r}
                   + \left[ g(\sigma) - g_s \phi_0 - 
                    \epsilon_k \right] u_k = 0 \; .
\end{equation}
The equation for the soliton field is
\begin{equation} \label{sigma_r}
    -\nabla^2 \sigma+U'(\sigma)
    + g'(\sigma) \rho_s(r) = 0.
\end{equation}
The quark density $\rho_q$ and the quark scalar density $\rho_s$
are given by
\begin{equation} \label{rho_q}
\rho_q(r) = \frac{n_q}{4 \pi {\bar k}^3 / 3 } \int_{0}^{\bar k} 
       d^3 k \; \left[ u_k^2(r) + v_k^2(r) \right] \; ,
\end{equation} 
\begin{equation} \label{rho_s}
\rho_s(r) = \frac{n_q}{4 \pi {\bar k}^3 / 3 } \int_{0}^{\bar k} 
       d^3 k \; \left[ u_k^2(r) - v_k^2(r) \right] \; ,
\end{equation} 
where the band is filled up to ${\bar k}$.
The quark functions are normalized to unity in the Wigner-Seitz cell.
The boundary conditions for the soliton field are 
$\sigma'(0)=\sigma'(R)=0$. The boundary conditions for the quark functions
at the origin are given by $u(0)=u_0$ and $v(0)=0$, where $u_0$ is determined
by the normalization condition
\begin{equation}\label{norm}
  \int_0^R 4 \pi r^2 d r (u(r)^2+v(r)^2) = 1 \; .
\end{equation} 
The boundary conditions at $r=R$ are given by
$u'_b(R)=0$ and $v_b(R)=0$ 
for the bottom of the lowest band, and
$u_t(R)=0$ 
for the top of this band.
Using these equations we can solve for the corresponding $\epsilon_b$
and $\epsilon_t$. We assume the tight-binding dispersion relation
$\epsilon_k = \epsilon_b^2+
(\epsilon_t-\epsilon_b)\sin^2(\pi s/2)$, with $s=k/k_t$,
and that the band is filled right 
to the top --- that is, ${\bar k}=k_t$. 
With this dispersion relation and filling, the nucleon
energy is given by
\begin{equation} \label{E_N}
  E_N = 3n_q \int_0^{1} ds \, s^2 
        \left\{\epsilon_b + (\epsilon_t-\epsilon_t)
       \sin^2\left({\pi s\over 2}\right)\right\}
      + \int_{0}^{R} 4 \pi r^2 d r 
        \left[ \frac{1}{2} \sigma'(r)^2 + U(\sigma) \right] \; .
\end{equation}
In order to correct for the spurious
center of mass motion in the Wigner-Seitz cell the nucleon
mass at rest is taken to be
\begin{equation} \label{M_N}
  M_N = \sqrt{E_N^2-<P_{cm}^2>_{WS}} \; ,
\end{equation}
where $<P_{cm}^2>_{WS}=n_q <p_q^2>_{WS}$ is the sum of
the expectation values of the squares of the momenta
of the $n_q$=3 quarks.

At low density the band width vanishes and the quarks are
confined in separate bags. Then we can assume the individual
nucleons move around as a gas of fermions with effective mass
$M_N$ given by Eq. (\ref{M_N}), so the nucleon energy
is $E^{(g)}_N=\sqrt{M_N^2+k^2}$. The
total energy density at nuclear density $\rho_B$ is thus
\begin{equation} \label{EeosG}
  {\cal E}_g = \frac{\gamma}{ (2 \pi)^3} \int_0^{k_F} 
      d \vec{k} \sqrt{M_N^2+k^2} + \frac{1}{2} m_s^2 \phi_0^2,
\end{equation}
where $\gamma=4$ is the spin-isospin degeneracy of the nucleons.
The Fermi momentum of the nucleons is related to the baryon density
through the relation
\begin{equation}\label{k_F}
	\rho_B=\frac{\gamma}{6 \pi^2} k_F^3 = {3\over 4\pi R^3} \; \; .
\end{equation}
The total energy per baryon is given by $E_g = {\cal E}_g / \rho_B $.
We have used the label $g$ to indicate that these expressions
correspond to a gas-like phase.
The constant scalar meson field $\phi_0$ is determined 
by the thermodynamic demand of minimizing ${\cal E}$:
\begin{equation} \label{minE}
  \frac{\partial {\cal E}_g}{\partial \phi_0} = 0 \; .
\end{equation}
Our mean field equations are similar to those of quantum hadrodynamics,
the difference here being that the nucleon now has structure and
thus the meson field couples to the nucleon through its quarks.

Let us review the treatment of dense matter in (I).
We started with the so-called Wigner-Seitz approximation, which
is often used in soliton calculations, since it is the simplest
picture of dense matter available. In this approximation, each
soliton is confined to a unit cell, which we can view from two
different perspectives. If we 
choose periodic conditions at the cell boundary, we have
the usual Bloch approach to the solid state: the quarks play
the role of the electrons and the scalar glueball field
takes the role of the ions in the usual crystal formulation.
If instead we
want to model the liquid state, we need to somehow introduce
some disorder into the system. From this view, then, we
take the Wigner-Seitz approximation as a sort of averaging over
the rest of the system: for the purposes of constructing an
individual nucleon, we ignore the motion of the nucleons,
instead adding ``by hand'' the kinetic energy of
a free gas to the system.
This is justified to the extent that
each nucleon's motion describes slow
degrees of freedom, whereas the constituents are fast degrees
of freedom that react essentially instantaneously to changes
in the relative arrangement of the nucleons --- that is, if
the Born-Oppenheimer approximation is valid.

Thus in (I) we calculated the equation of state of nuclear
matter within the second scheme. The mass of
the nucleon was calculated
as a function of the radius of the (spherical)
Wigner-Seitz cell, for which it was then necessary to subtract
away spurious center-of-mass motion. This was done using
approximate relations discussed in detail in \cite{Wil89}.
Note that within the first scheme, the quarks are not
assumed to be in a state of good momentum, and this kinetic
energy is not spurious.

\section{Significant Structure Theory}
\label{sec:liquid}

In (I) we used the Wigner-Seitz approximation to determine the
effective nucleon mass in nuclear matter, but then gave the
nucleons the kinetic energy of a Fermi gas. In this approximation,
the quarks feel the nuclear medium, but the nucleons themselves
do not.  That is, we assume the quarks adjust instantly to the medium, 
forming 3$q$ collective states that move relatively slowly 
and essentially freely through the medium.
The interactions between nucleons occur only indirectly
through the effective mass and the mesonic mean field. Clearly,
such an approximation cannot be accurate at high densities, when
the finite size of the nucleon becomes important, for
nucleons will not then move freely. Instead, we need to model the
liquid state, where any individual nucleon will range about the
system over long time scales, but will be localized on shorter
time scales. One would still like to approach the problem using the
Wigner-Seitz approximation, insisting now that the nucleon also
feels the medium and does not leave the cell.
Clearly, this is a very restrictive assumption, and physical
chemists long ago understood that this corresponds more closely to
the solid than the liquid state. Nevertheless, this will
provide a starting point for the model of the liquid state we
shall adopt in the following, so let us pursue this approach
further.

Ideally, we would like to allow our nucleon to rattle about
in its Wigner-Seitz cell in order to extract a potential.
This would entail dropping the assumptions of spherical symmetry
for the bag and the use of only $s$-wave quark states. Instead, we
shall attempt to model the nucleon's motion at high density
as follows. First, we assume that the motion is harmonic --- that
is, the center of mass ${\bf R}_{cm}$ of the nucleon is never far from the
center of the WS cell ${\bf R}_0=0$.
Now, the average of a harmonic potential
in the ground state is 
\begin{equation}
\langle V \rangle =
\left\langle {1\over 2}M_N\omega^2_N {\bf R}_{cm}^2 
\right\rangle = {3\over 4}\omega_N,
\end{equation}
from which we find the natural frequency $\omega_N$ as a function
of the effective nucleon mass $M_N$ and the mean square of the
center of mass coordinate. 
Next, we identify ${\bf R}_{cm}$ as
the center of the quark distribution in the cell, ignoring thereby
any motion of the soliton bag $\sigma$. 
(Viewing Fig. (3) of (I), we see that the soliton bag
begins to be ``squeezed'' by the WS boundary at $R\approx 1.5$fm,
so that at least for densities higher than this we might argue
that the bag is essentially fixed.)
Taking ${\bf R}_{cm}$ to be the
center of mass coordinate of three quarks, we have
\begin{equation}
\langle {\bf R}_{cm}^2 \rangle = \left\langle\left[ {1\over 3}
({\bf r}_1 + {\bf r}_2 + {\bf r}_3)\right]^2\right\rangle
= {1\over 3} \langle r_q^2 \rangle_{WS}.
\end{equation}
Now we identify the last average with the mean square charge radius
of the nucleon
\begin{eqnarray}
\langle r^2 \rangle_{WS} &=& 
\frac{\int_0^R d^3r\, r^2 \rho_q(r)}{\int_0^R d^3r\, \rho_q(r)} \\
\nonumber
&=& {1\over {4\over 3}\pi {\bar k}^3} \int_0^{\bar k} d^3k \,\,
\frac{\int_0^R d^3r\, r^2 \left[u^2_k(r)+v^2_k(r)\right]}
{\int_0^R d^3r \left[u^2_k(r)+v^2_k(r)\right]}.
\end{eqnarray}
This is clearly only a rough estimate of the center of mass
motion of the nucleon. The nucleon energy in the solid is now
\begin{equation}\label{EsN}
E^{(s)}_N = M_N + {3\over 2} \omega_N, \quad {\rm with} \quad 
\omega_N = \frac{3}{2 M_N \langle R_{cm}^2\rangle}.
\end{equation}
This approximation corresponds to subtracting away spurious kinetic energy
from the soliton energy Eq. (\ref{E_N}), namely,
\begin{eqnarray}
E_{sp} = E_N - E^{(s)}_N &=& 
\sqrt{M_N^2 + \langle P^2_{cm}\rangle_{WS}}
- M_N - {3\over 2}\omega \\ \nonumber
&\approx& {1\over 2M_N}\left(
\langle P^2_{cm}\rangle_{WS} - {9\over \langle 2R_{cm}^2 \rangle_{WS}}
\right).
\end{eqnarray}
Approximating $\langle R_{cm}^2 \rangle$ by 
$\frac{1}{3}\langle r^2\rangle_{WS}$ can only be valid at high density.
At low density, this surely breaks down for then the quarks
cannot reach the WS boundary unless the bag itself is allowed
to move. Moreover, the present approximation corresponds to
treating the soliton matter as an Einstein solid (we have implicitly
averaged over the Bloch momenta ${\bf K}$ corresponding to the lattice
of nucleons: a more accurate treatment 
of the solid would put the three quarks
in a state of good ${\bf K}={\bf k}_1+{\bf k}_2+{\bf k}_3$ 
and find a frequency $\omega_N({\bf K})$
that is a function of the Bloch momentum).
The total energy density in this ``solid'' phase is
\begin{equation} \label{EeosS}
  {\cal E}_s = \rho_B \left( M_N + \frac{3}{2} \omega_N \right)
       + \frac{1}{2} m_s^2 \phi_0^2.  
\end{equation}
The constant scalar meson field $\phi_0$ is again determined
by the thermodynamic demand of minimizing ${\cal E}$.
Both $M_N$ and $\omega$ depend implicitly upon $\phi_0$, so
that this equation must be solved iteratively in conjunction
with those for $\sigma$ and $\psi$.

This gives the equation of state for solid nuclear matter, which is
not of much interest in itself.
However, this is useful for building a model of the liquid
state at high density based upon significant structure
theory \cite{JE69}. The essential idea is to isolate those
configuations that make the significant contribution to
the partition function. Based upon experimental observations
of molecular liquids, the liquid state is viewed as a close-packed
lattice with holes present that
destroy any long-range order. The volume of the system increases
by increasing the number of holes, with the volume of each hole equal
to the average volume occupied by a close-packed molecule, since
this balances the competing demands for greater entropy and
lower energy. A molecule neighboring a hole can move into
the vacancy, creating a new hole at the site it left. Each hole
thus replaces three vibrational with three
translational degrees of freedom. Thus the liquid is
represented as a combination of molecules with solid-like
properties (those next to filled sites) and gas-like properties
(those next to vacant sites).

Now consider this model of the liquid state applied to dense
solitonic matter. (Such an application has been studied previously
for dense skyrmion matter in Ref. \cite{Wal90}.)
First, we note that our assumptions of a spherical Wigner-Seitz
cell, which reflects a sort of average over nearest neighbor
positions, and the Einstein approximation that $\omega_N$ is
independent of the Bloch momentum ${\bf K}$, which ignores long-range
correlated vibrations of the nucleons --- approximations that
would be rather severe if we truly wished to model a solid ---
are instead appropriate for the present application. We need
only add the holes to ensure the disorder corresponding to
a liquid state. So let $V_l$ be the total volume of the liquid and
$v$ be the volume of each cell (occupied or unoccupied). If there
are $N$ nucleons and $N_h$ holes, then $V_s = Nv$ is the total
volume of the occupied cells and $V_g = V_l - V_s = N_h v$ is
the total volume of the holes. On average, a nucleon will
encounter a neighbor on a fraction ${N\over N+N_h}={V_s\over V_l}$
of its trips and a hole on a fraction
$1-\frac{V_s}{V_l}=\frac{V_g}{V_l}$ of its trips. Thus there
are $3N\frac{V_s}{V_l}$ solid-like and $3N\frac{V_g}{V_l}$ gas-like
degrees of freedom, and the liquid partition function
is
\begin{equation}
Z_l = Z_s^{N{V_s\over V_l}} Z_g^{N(1-{V_s\over V_l})}.
\end{equation}
This results in the following nucleon energy in the liquid
state:
\begin{equation}
E_N^{(l)} = {V_s\over V_l} E_N^{(s)}(v)
+{V_l-V_s\over V_l} \tilde{E}_N^{(g)}(n_g),
\end{equation}
where $n_g = \frac{N_g}{V_g} = \frac{N}{V_l}$ is the ``density''
of the gas-like part of the system, which from the last equality
is equal to the baryon density $\rho_B$. The average
solid-like nucleon energy  $E_N^{(s)}(v)$ is given by
Eq. (\ref{EsN}), with $M_N$ and $\omega_N$ depending upon
$v$ through the Wigner-Seitz radius $R=(3v/4\pi)^{1/3}$.
The average gas-like nucleon energy is instead
\begin{equation}\label{Eg}
\tilde{E}_N^{(g)}(n_g) = {3\gamma\over 4\pi k_g^3}
\int^{k_g}_0 d^3k \sqrt{M_N^2(v_l) + k^2},
\end{equation}
with $k_g = (6\pi^2 n_g/\gamma)^{1/3}
= (6\pi^2\rho_B/\gamma)^{1/3}$ and $v_l=V_l/N=1/n_g$. The total energy
density of the system in the liquid phase is then
\begin{equation}\label{EeosL}
{\cal E}_l (v) = \rho_B^2 v E_N^{(s)}(v) +
\rho_B(1-\rho_B v)\tilde{E}^{(g)}_N(\rho_B)
+{1\over 2}m_s^2\phi_0^2 - {1\over 2}m_V^2 V_0^2.
\end{equation}
Once again, $\phi_0$ is determined by minimizing ${\cal E}_l$,
and so for nonzero $g_s$ we must solve anew the equations for
$\sigma$ and $\psi$ consistent with the mean field value 
$\phi_0(v,\rho_B)$
determined from this new liquid equation of state.
In Eq. (\ref{EeosL}), the cell volume $v$ is to be taken
as a parameter. Note that the WS cell volume is no longer
$1/\rho_B$ when holes are present. Instead, we can define
a new radius $R_l = (3/4\pi\rho_B)^{1/3}$ which is half the
average spacing between nucleons in the liquid state.

There is some question as to whether one should take the
effective nucleon mass in (\ref{Eg}) to be $M_N(v)$ or
$M_N(v_l)=M_N(\rho_B^{-1})$. Since a single
nucleon can have both gas-like and solid-like degrees of
freedom, it might be more consistent to use the former
choice in calculating the gas-like part of the nucleon
energy. However, there seems little reason to insist that
the inertial masses corresponding to the solid-like harmonic
motion and the gas-like translational motion are the same.
Moreover, the latter choice guarantees that our liquid EOS
reduces to the Fermi gas EOS given by (\ref{EeosG}) at
low density. Thus our equation of state (\ref{EeosL}) interpolates
smoothly between high-density and low-density
pictures of the liquid state.

\section{Results}
\label{sec:results}
\subsection{Nuclear matter}

In Figs. 1-3 we show the energy per baryon of $\chi$CD solitonic
nuclear matter for three different values of the quark-meson
coupling. In each graph we display the ``gas'' equation of
state (\ref{EeosG}) and a set of 
significant liquid structure curves characterized by
various choices of the WS cell volume $v=4\pi R^3$ in
(\ref{EeosL}). In addition, the upper curve labelled ``solid''
in these figures is given by (\ref{EeosG}) with the
condition $n_g = 1/v$ --- or, equivalently, $R_l=R$ --- which
implies that the number of holes is zero, so that the nuclear
motion is entirely solid-like. Note that our approximations
in estimating $\omega_N$ break down at low density. In particular,
the ``solid'' curve is not to be trusted above $R\approx 1.5$fm:
here, the bag is no longer squeezed by the WS boundary and its
motion probably can no longer be ignored.

As $v\to0$, the significant structure curve approaches
the ``gas'' curve, which certainly underestimates the energy per
nucleon. Without empirical evidence of a transition from liquid
to solid nuclear matter, we are unable to fix $v$ directly. (Some
models predict such a transition; however, it is more widely
believed that the transition to the quark-gluon plasma will
preclude a transition to solid nuclear matter.) Here, we can
just treat $v$ as a parameter in choosing the EOS that best
fits the empirical data at the saturation point. We find that,
for the parameter sets we studied, the curves that best fit 
the empirical EOS for nuclear matter are given by the values
$g_s = 2, R=0.4$fm and $g_s = 3, R=0.7$fm. The values of
the Fermi momentum, binding energy, and compression modulus 
at the saturation point are $k_s = 1.06$fm$^{-1}$,
$E_s= 22$MeV and $K = 1082$MeV for the
former, and $k_s = 0.95$fm$^{-1}$, 
$E_s= 24$MeV and $K = 671$MeV 
for the the latter curve.  These are to be compared with
the empirical values $k_s = 1.36$fm$^{-1}$, 
$E_s= 16$MeV and $K \approx 200$MeV. 
In particular, we see a significant improvement in the 
value of the compression modulus with respect to the
calculation of (I), although it is still rather high with
respect to empirical values.

\subsection{Quark matter}

In the high density limit, the preferred phase in the nontopological
soliton models we are studying is a uniform plasma characterized
by the solution $\sigma=0$. That is, the soliton bags disappear,
and one is left with a quark gas. 
The energy density of the quark plasma is
\begin{equation}\label{Qeos}
{\cal E}_q = {3 k_F^4\over 2\pi^2} + B,
\end{equation}
where the Fermi momentum is related to the baryon density $\rho_B$
as $k_F=(6\pi^2\rho_B/\gamma)^{1/3}$ --- the degeneracy of the
quark gas is $n_q\gamma$ and the quark density is $n_q$=3 times
the baryon density. The bag constant is $B=46.6$$MeV/fm^3$ for
our choice of parameters. 
In contrast to our assumption about the solitonic phase,
the energy density of the quark gas is altered significantly
by perturbative gluonic contributions, especially at higher densities.
In this phase, the $\chi$CD model in the mean field approximation
is equivalent to perturbative QCD.
As shown in \cite{FM77}, for example,
adding the lowest order gluonic corrections then gives
\begin{equation}\label{QGeos}
{\cal E}_q = {3 k_F^4\over 2\pi^2}\left\{1 + {2\alpha_s\over 3\pi}
+{\alpha_s^3\over 3\pi^2}
\left[6.79+2\ln\left({2\alpha_s\over\pi}\right)\right]\right\} + B,
\end{equation}
where the leading-logarithm expression for the strong coupling
constant is
\begin{equation}\label{alpha_s}
\alpha_s(k_F) = \frac{6\pi}{29\ln(k_F/\Lambda)}.
\end{equation}
We take the QCD scale parameter to be $\Lambda\sim 180$-$200$MeV.

In Fig. 4 we show the two nuclear matter curves selected above
along with the quark-gluon plasma EOS given here. In addition,
we show an ``empirical'' nuclear matter EOS given by
\begin{equation} \label{EmpEOS}
E_B \approx {K\over 18}\left( {k_F^3\over k_s^3} - 1\right)^2
+ M^{(as)}_N - E_s,
\end{equation}
with $M^{(as)}_N-E_S = 1160$MeV and $K=200$MeV. (The bag constant
$B$ appearing in the quark-gluon plasma energy has been set in fitting
the free soliton mass and rms radius, and thus for purposes
of comparison we take the
low density limit of the ``empirical'' curve to coincide with the
free soliton mass.) Note that the $g_s=2$ shown here has a transition to 
the solid state at $k_F=(9\pi)^{1/3}/2R=3.8$fm$^{-1}$ and the
$g_s=3$ at $2.2$fm$^{-1}$, both well past the transition points
to the quark-gluon plasma. Clearly, the saturation points of the
model curves occur at densities that are too low. Nevertheless, the
very fact that we find qualitative agreement with the empirical
nuclear matter EOS is quite encouraging.

\section{Conclusions and Outlook}
\label{sec:con}

In this paper we have developed an improved modeling of the
liquid state of solitonic matter based upon the significant
structure model used in physical chemistry. We have applied
this to the study of nuclear matter within
a nontopological soliton model with explicit quark degrees
of freedom. In particular, initial studies in (I) indicated that
among several such similar models,
the chiral chromodielectric model gave results most in line
with empirical expectations. When a scalar meson is allowed
to couple to the quarks, saturation can be achieved.
The calculations in (I) were based upon
a modeling of nuclear matter that assumes the solitons move
essentially freely through the medium. This clearly
will break down at and above nuclear saturation density,
where the size of the individual solitons and the spacing
between solitons in the medium become comparable.
The significant structure
model of the liquid state used in the present paper, however, is
designed for densities near the transition to the solid
state. We have thus developed here an equation of state that
interpolates between models designed for low and high
densities. It is encouraging to find that without changing
the free nucleon properties we can adjust the quark-meson
coupling to reproduce the saturation point of nuclear matter
found in (I). Indeed, we find that the compression modulus
is improved significantly when calculated using the significant
liquid structure model.

With the $\chi$CD model we are able to treat the transition
to the quark-gluon plasma consistently. This is the point of
developing this model: confinement is dynamical. Thus the
parameters governing the nuclear and plasma phases are
in principle the same. Clearly, the results we have found
here, while qualitatively in agreement with empirical
estimates, are not quantitatively useful. In particular,
the saturation point of nuclear matter occurs at far too
low a density for the parameters used in our calculation.
Moreover, the mass of the free nucleon is too high. We
have not tried too hard to improve our results by adjusting
the parameters of the model. Probably one can find a better
choice of parameters than those for which we have done the calculations
here, but we must emphasize that there seems to be a limit
in how well one can do. This can be seen already in studying
the free nucleon. There, we are unable to find a parameter
set that reproduces the nucleon mass and rms radius exactly while
still giving reasonable values for the glueball and bag constant.
We compromised here by fitting the nucleon rms radius well
while keeping the glueball mass and bag constant in their
accepted ranges. This resulted in a free nucleon mass that was too high.

In trying to reproduce the empirical saturation point of
nuclear matter, however, we found a general scaling
phenomenon. Keeping the saturation energy roughly correct, it
seems difficult to change the parameters in such a way as to
get the correct density expect by a rough overall scaling
of all predicted quantities. Thus getting the correct saturation
density entails lowering the rms radius along with a corresponding
increase in the mass of the nucleon. This is suggestive. As a
matter of fact, the quark-meson model we developed here is
better suited for nuclear matter than for isolated nucleons.
This is because an isolated nucleon will surround itself with
a pion cloud, whereas (unless pion condensation occurs) the
effects of pions in nuclear matter are likely to be small if
we already have scalar mesons. Thus we can consider the model used
here as actually better used for describing the quark core
of the nucleon. The addition of the pion will allow
the quark core to shrink and lower the energy of the
free nucleon with respect to that of the quarks. With the shrinking
of the quark core, one can expect a corresponding decrease in
the volume per nucleon at the saturation point.

Thus we view the addition of pions as an essential improvement
to our model. This, of course, was always obvious: we certainly
cannot hope to reproduce the long-range part of the
nucleon-nucleon interaction without pions. Our results
indicate that the presence of pions is necessary in order 
to reproduce the structure of the nucleon as well. This
comes as little surprise. 
There are of course other improvements that can be made upon
our calculations, such as a better handling of the
Bloch boundary conditions and quark wave functions \cite{Web97}
and an improved treatment of the corrections due to spurious
center of mass motion. Perturbative
corrections due to gluons and mesons about the mean field should
eventually be considered as well. 

Having said (that is, written) this about improving the
model, we should not lose sight of the original goal of our
work. What we wanted to do first of all was see if we could distinguish
among the various nontopological soliton models on the
market by studying dense matter. To this we can answer that the
chiral chromodielectric model appears to be more in line with
empirical expectations. Then, we wanted to develop as simple a model
as possible that could give a rough qualitative fit to both
free nucleon and dense nuclear matter properties, thus providing
a reasonable starting point for more sophisticated models that
can provide truly quantitative predictions. To this we can
also answer that the chiral chromodielectric model, when modified according
to the local uniform approximation by the addition of a scalar
meson field, would appear to be such a model. In fact, this
model not only gives a good qualitative and rough quantitative
fit to the empirical equation of state, it also predicts an
increase of the nuclear rms radius in the nuclear medium, a
result that is in accord with the EMC effect. Moreover, the
results presented here seem to indicate that with the addition
of pions we would have good possibilities of obtaining a quantitatively
accurate fit to single nucleon properties and to the
empirical nuclear matter equation of state. This would provide
a reliable estimate of the bag constant and therefore
a consistent and accurate treatment of the transition from nuclear
to quark-gluon matter as 
a true transition between two phases of a single model.


\newpage

\begin{figure}[htb]
\caption{\protect\label{fig1}
Energy per baryon as a function of average liquid radius
$R_l=(3/4\pi\rho_B)^{1/3}$ for various values of the
Wigner-Seitz radius $R=(3v/4\pi)^{1/3}$ for quark-meson
coupling $g_s=1$. Also shown are the curves for the
``gas'' limit $R\to 0$ and the ``solid'' limit $R_l = R$.}
\vskip-2.2cm
\hskip-1.5cm
\includegraphics{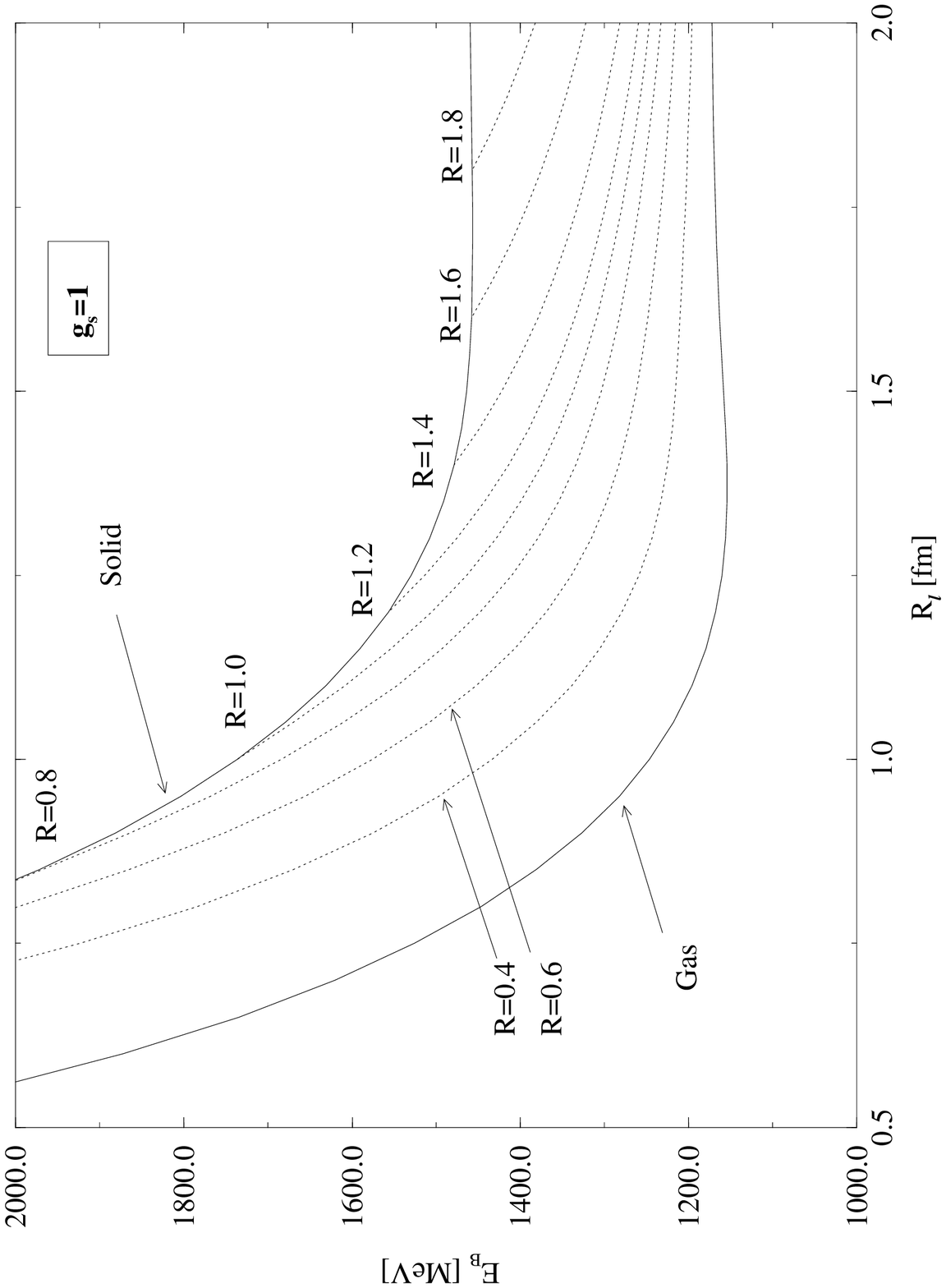}
\vskip8.0cm
\end{figure}

\,
\newpage

\begin{figure}[htp]
\caption{\protect\label{fig2}
As in Fig. 1, for the choice of quark-meson
coupling $g_s=2$.}
\vskip-2.5cm
\hskip-1.5cm 
\includegraphics{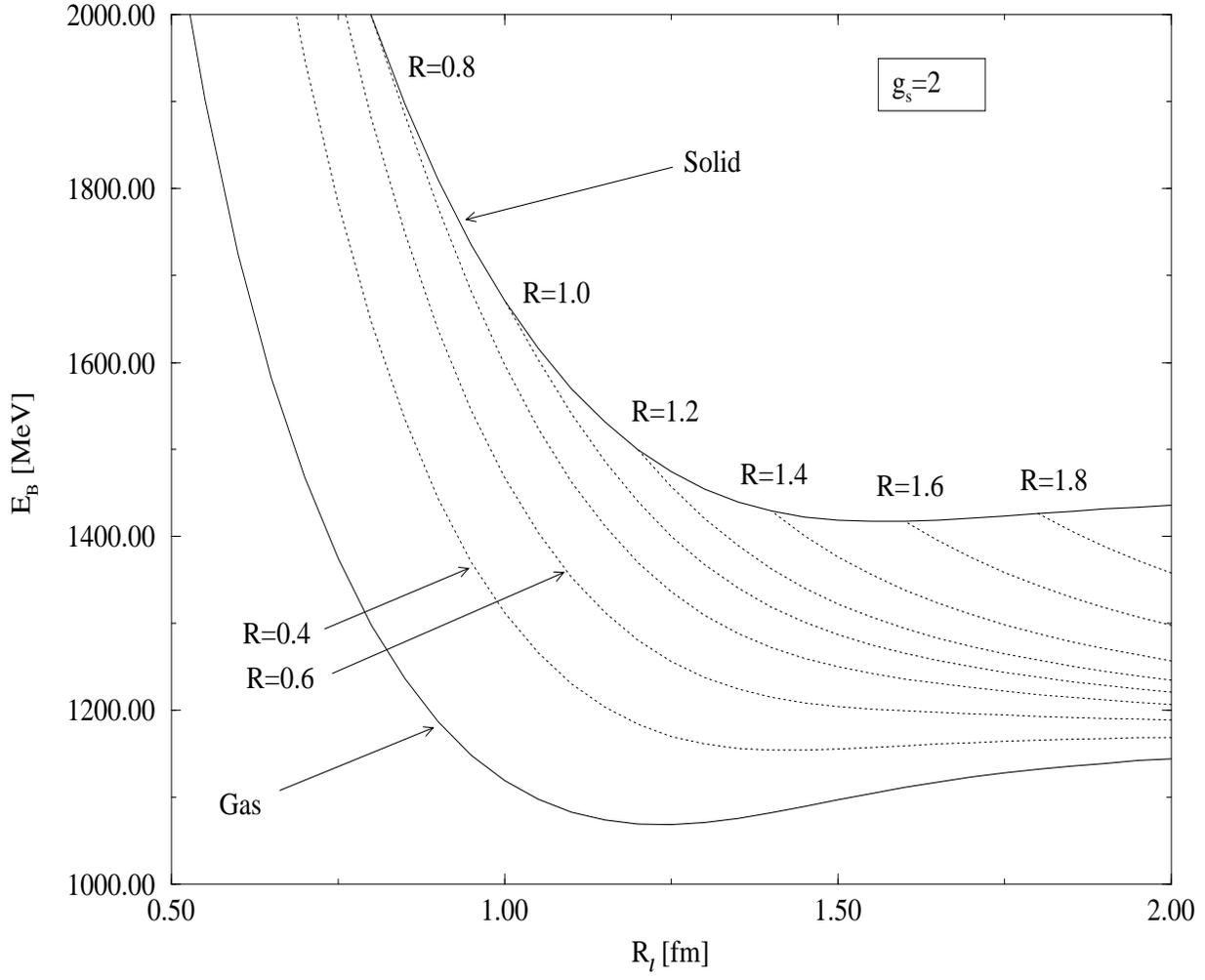} 
\vskip4.0cm
\end{figure}

\newpage

\begin{figure}[htp]
\caption{\protect\label{fig3}
As in Figs. 1 and 2, but with $g_s=3$.}
\vskip-2.5cm
\hskip-1.5cm
\includegraphics{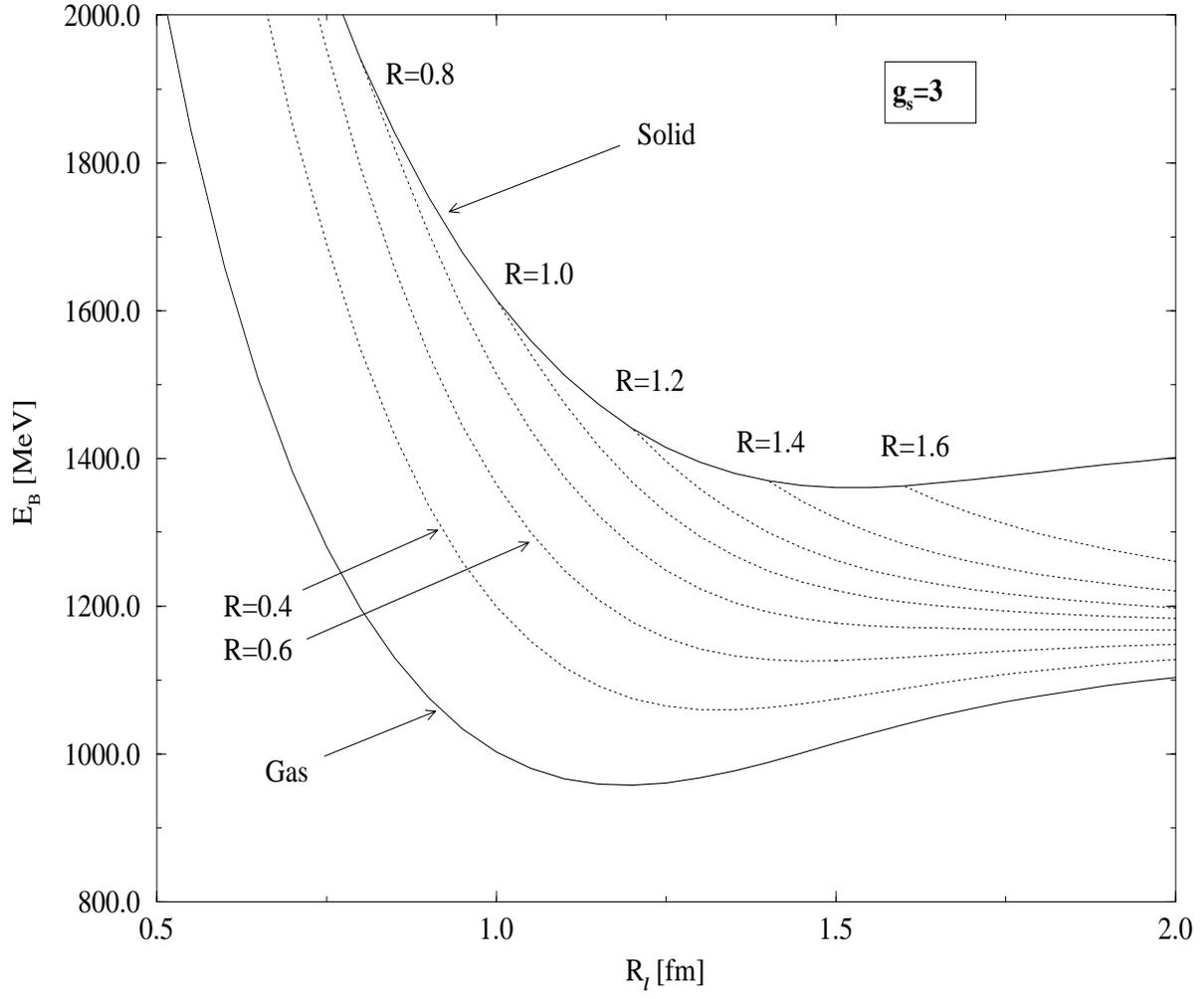}
\vskip4.0cm
\end{figure}

\newpage

\begin{figure}[htp]
\caption{\protect\label{fig4}
Quark-gluon plasma (QGP) and nuclear matter energy per baryon number
as function of the Fermi momentum. The curves are as
described in the text.}
\vskip-2.3cm 
\hskip-1.0cm 
\includegraphics{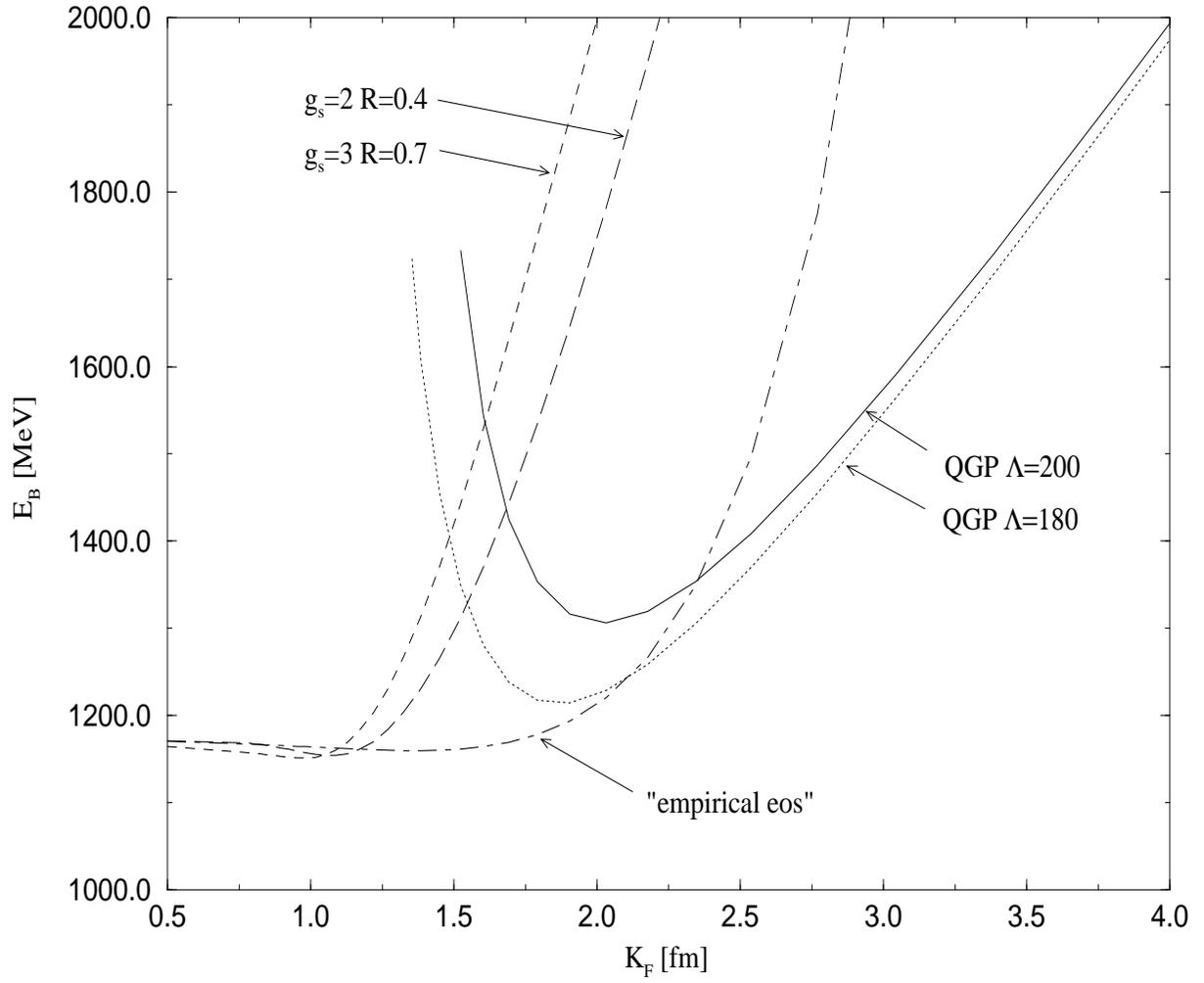} 
\vskip4.0cm
\end{figure}

\end{document}